\documentclass[toc,cits]{PoS}

\usepackage{amssymb,fontenc,times,mathptmx,graphicx}

\title{CMB interferometry}

\ShortTitle{CMB interferometry}

\author{\speaker{Clive Dickinson}\thanks{CD acknowledges an STFC Advanced Fellowship and an EU Marie Curie IRG grant under the FP7.}\\
        Jodrell Bank Centre for Astrophysics, Alan Turing Building, School of Physics \& Astronomy, The University of Manchester, Oxford Road, Manchester, M13 9PL, U.K.\\
        E-mail: \email{Clive.Dickinson@manchester.ac.uk}}

\abstract{Interferometry has been a very successful tool for measuring anisotropies in the cosmic microwave background. Interferometers provided the first constraints on CMB anisotropies on small angular scales ($\ell \sim 10000$) in the 1980s and then in the late 1990s and early 2000s made ground-breaking measurements of the CMB power spectrum at intermediate and small angular scales covering the $\ell$-range $\approx 100$--$4000$. In 2002 the DASI made the first detection of CMB polarization which remains a major goal for current and future CMB experiments. Interferometers have also made major contributions to the detection and surveying of the Sunyaev-Zel'dovich (SZ) effect in galaxy clusters. 

In this short review I cover the key aspects that made interferometry well-suited to CMB measurements and summarise some of the central observations that have been made. I look to the future and in particular to HI intensity mapping at high redshifts that could make use of the advantages of interferometry.}

\FullConference{Resolving the Sky - Radio Interferometry: Past, Present and Future -RTS2012\\
		April 17-20, 2012\\
		Manchester, UK}

\newcommand{\mnras}{MNRAS}
\newcommand{\apj}{ApJ}
\newcommand{\apjl}{ApJL}

\newcommand{\nat}{Nature}
\newcommand{\pasp}{PASP}
\newcommand{\nar}{New Astronomy Reviews}

\begin{document}

\section{CMB interferometry}

\subsection{Interferometry basics}

An interferometer measures the complex correlation between pairs of antennas. In the flat-sky approximation, the observed ``visibilities'', $V(u,v)$, are samples of the $u,v$-plane which corresponds to the Fourier Transform (FT) of the sky intensity distribution, $I(x,y)$, given by the van Cittert-Zernike equation:

\begin{equation}
V(u,v) = \int du dv A(x,y) I(x,y) e^{2\pi i(ux + vy)}
\end{equation} 

where $A(x,y)$ is the primary beam response of the antennae. Essentially, the visibilities are samples of the FT of sky distribution multiplied by the primary beam function. The visibilities can then be inverted to produce images of the sky distribution convolved with the synthesized beam. 

The primary cosmological quantity of interest is the CMB power spectrum, $C_\ell$, since it is known that the CMB anisotropies are very close to Gaussian. The power spectrum is essentially the FT (or more stricty, the spherical transform) of the sky, and thus interferometers measure the power spectrum almost directly. Essentially, averaging the square of the visibilities, $|V({\bf u})|^2$, in shells of constant ${\bf u}$, gives a (noise-biased) estimate of $C_{\ell}$  at $\ell=2\pi {\bf u}$, convolved with the window function $W$. This is accurate in the flat-field approximation (typically $\ell \gtrsim 100$). The multipole window function is determined by the FT of the primary beam (the aperture illumination function), although the $\ell$ resolution can be improved by mosaicing \cite{Myers03}.

\subsection{Advantages of CMB interferometry}

Interferometers have a number of key advantages compared to total-power experiments; these are summarised in Table~\ref{tab:advantages}. Interferometers are inherently more stable than total-power receivers since they measure the correlation of signal between pairs and thus are not dependent on the total-power (baseline) response of individual detectors. Interferometers do not measure the DC signal, such as the 2.7\,K CMB temperature and the bulk of atmospheric emission, which is a major advantage. Long time-scale drifts of receivers ($1/f$ noise), which is a major problem for total-power experiments is also mitigated by implementing Automatic Gain Correction (AGC) in the IF before correlation. Phase-switching naturally provides further suppression of systematics. This allows the thermal noise to be realised even for very long (months and years) integration times. 

Another key advantage is the attenuation of non-astronomical signals, which have different fringe rates to signals from within the field-of-view\footnote{Strictly, the DASI/CBI interferometers could not utilise fringe-rate filtering due to their non-tracking antennas, and had to use differencing techniques to remove local correlated signals.}. In fact, fringe-rate filtering can be put to good use for filtering out known sources including Sun, Moon and other local signals (see \cite{Watson03} for a demonstration). Accurate calibration is crucial for measuring the faint CMB fluctuations, particularly in the presence of bright foreground sources (radio galaxies, diffuse foregrounds etc.). Calibration of interferometric data is well-studied and has been shown to be capable of producing images with enormous dynamic ranges. Furthermore, the synthesised beam is known to high precision since it depends only on the $u,v$ sampling which is known to high accuracy.

\subsection{Disadvantages of CMB interferometry}

There are however some drawbacks, as summarised in Table~\ref{tab:advantages}. For measuring diffuse (extended) sources, such as the CMB, it is the surface brightness sensitivity (e.g. in units Kelvin, or Jy/unit area) that matters, rather than point source sensitivity (in units Jy). Therefore, the size of the dishes is not the critical quantity, but rather the filling factor, $f$, of the array, as defined by
\begin{equation}
f = N (D/d)^2
\end{equation}
where $N$ is the number of antennas, $D$ is the dish diameter, and $d$ is the longest baseline. The relation between brightness temperature $T_b$ and flux density $S$ is then given approximately by 
\begin{equation}
S = \frac{2kT_b \Omega}{f \lambda^2} 
\end{equation}
For a single dish, $f \approx 1$ and is the optimal that can be achieved. This is one of the principal drawbacks for interferometers, and this is why total-power (single dish) experiments with large detector arrays are now preferred. The other drawback is in measuring the largest angular scales, which are inherently difficult with an interferometer. They require very short baselines, which results in shadowing\footnote{DASI and CBI used co-mounted dishes which did not suffer from the effects of shadowing and achieved very high filling factors ($f \approx 1$)} and the largest angular scales (larger than the primary beam) are essentially impossible to measure. For the same reasons, aperture synthesis maps are difficult to quantify on scales $>>$ synthesised beam because of resolution effects i.e. loss of flux on large angular scales.

\begin{table}[!t]
\small
\begin{tabular}{c}
\hline
Accurate beam knowledge ($u,v$ coverage known exactly)    \\
Accurate window function (FT of primary beam) \\
Rejection of DC signals \\
Strong rejection of systematics (baselines, rejection of non-astronomical signals, no $1/f$ noise, pointing) \\
Direct access to power spectrum (square of visibilities) \\
High dynamic range \\
Maps via synthesis imaging \\
Easy access to wide range of angular scales (array configuration) \\
\hline
Not as sensitive as total-power (filling factor) \\
Largest angular scales difficult (shortest spacing) \\
Maps are difficult to quantify (flux loss on large angular scales/resolving out) \\
\hline
\end{tabular}
\caption{A list of some of the advantages ({\it top}) and disadvantages ({\it bottom}) of interferometry for CMB analyses compared to total-power experiments.}
\label{tab:advantages}
\end{table}
\normalsize

\section{Summary of observations}
\label{sec:obs}

There have been a number of early attempts of detecting CMB anisotropies with interferometers going back to the early 1980s\footnote{The author knows of early attempts by Prof Rodney Davies using the Lovell Mk1 and Mk2 dishes at Jodrell Bank but these were never published due to lack of sensitivity.}. A summary of interferometric CMB measurements is given in Table~\ref{tab:obs}.

\subsection{Early measurements at high resolution}

The first published work that the author is aware of was with the newly commissioned Very Large Array (VLA) by \cite{Fomalont84}. A number of similar measurements were made using the OVRO, Ryle, BIMA and ATCA arrays (see Table~\ref{tab:obs}). These gave constraints on the very high $\ell$ spectrum and on the contribution from unresolved point sources. 

The first dedicated CMB array was the Cambridge Anisotropy Telescope (CAT). Built by Cambridge University at Lord's Bridge, it consisted of 3 antennas located inside a ground screen and operated at 15\,GHz. It was the first interferometer to detect primary CMB anisotropies at $\ell \sim 500$ \cite{Scott96}. In many ways the CAT experiment paved the way for the more sensitive arrays in the next decade. A dedicated single baseline interferometer was also employed in Tenerife, operating at 33\,GHz, and managed to detect anisotropies on $1^{\circ}$ and $2^{\circ}$ scales \cite{Harrison00}.

\begin{table}[!t]
\small
\begin{tabular}{lccccccc}
\hline
         &         &              &Frequency   &Bandwidth   &                &           &Reference \\
  Name   &Location &$N_{\rm dish}$  &(GHz)      &(GHz)        &Primary beam   &$\ell$     &           \\
\hline
OVRO     &U.S.     &6             &30         &2.0          &$4^{\prime}$        &6750   &\cite{Carlstrom96}  \\
VLA      &U.S.     &27            &8          &0.2          &$5^{\prime}$        &6000   &\cite{Fomalont84,Partridge97}  \\
IRAM     &France   &3             &88         &0.4          &$55^{\prime\prime}$  &3000--70000 &\cite{Radford93} \\
Ryle     &England  &8             &15         &0.4          &$6^{\prime}$        &4500   & \cite{Jones93} \\
BIMA     &U.S.     &10            &30         &0.8          &$6^{\prime}$        &5000--9000   & \cite{Dawson02} \\
ATCA     &Australia &6            &9          &0.1          &$8^{\prime}$        &3400   & \cite{Subrahmanyan98} \\
T-W      &U.S.     &2             &43         &...          &$2^{\circ}$          &20--100 &  \cite{Timbie90} \\
IAC-Int  &Tenerife &2             &33         &0.5          &$2^{\circ}$           &110--220        & \cite{Harrison00} \\
CAT      &England  &3             &13--17     &0.5          &$2^{\circ}$          &339--722 &  \cite{Scott96} \\
VSA      &Tenerife &14            &26--36     &1.5          &$4.\!^{\circ}6/2.\!^{\circ}1$  &150--1600 & \cite{Dickinson04} \\
DASI     &South Pole &13          &26--36     &10.0         &$3^{\circ}$          &125--700  &  \cite{Halverson02,Kovac02}\\
CBI      &Chile    &13            &26--36     &10.0         &$45^{\prime}$/$28^{\prime}$ &630--3500 &  \cite{Padin02,Readhead04b,Taylor11} \\
SZA      &U.S. &8      &30 \& 90   &8            &$12^{\prime}/4^{\prime}$     &4000      &\cite{Sharp10,Muchovej11} \\
\hline
\end{tabular}
\caption{Summary of interferometric experiments to measure CMB anisotropies. This is an update of a previous version of this table produced by \cite{White99}.}
\label{tab:obs}
\end{table}
\normalsize

\subsection{The era of the CMB interferometers}

In the late 1990s, three semi-independent collaborations began designing much more sensitive and flexible arrays dedicated to measuring the CMB power spectrum. The Cambridge, Jodrell Bank and IAC groups worked together to construct the Very Small Array (VSA). In the U.S., the Caltech group formed two independent collaborations to build the Degree Scale Instrument (DASI) and the Cosmic Background Imager (CBI). These three instruments were pivotal in making the first precision measurements of the CMB power spectrum on sub-degree scales.

The VSA was a 30\,GHz 14-element interferometer that could be reconfigured to probe angular scales from $\sim 10^{\prime}$--$2^{\circ}$. It was originally designed to have several 1.5\,GHz bands but funding was only given to utilise a single 1.5\,GHz channel. The VSA was carefully designed to minimise all systematics, both instrumental and astrophysical. The entire telescope was located inside a large ground-screen while the tracking elements on the tip-tilt table allowed fringe-rate filtering of non-astronomical signals such as cross-talk and and ground spillover \cite{Watson03}. This proved to be essential for observing during the day (to filter out the Sun and Moon) and also for filtering out correlated signals that were found to be strong on the shortest baselines. The VSA also employed an additional single-baseline interferometer consisting of two 3.7\,m dishes located in their own ground screens. This allowed for a survey of the brightest sources in each VSA field (positions were provided from an earlier 15\,GHz survey of the VSA fields by the Ryle telescope) at the same frequency and at the same time. The VSA produced accurate measurements of the CMB power spectrum over the range $\ell=150$--$1600$ \cite{Dickinson04}.

The DASI instrument was located at the South Pole and used 13 20\,cm horns located on a table mounted onto an az-el mount. It operated at 26--36\,GHz with 10 channel 10\,GHz digital correlator. Its small horns and table provided good sensitivity over the $\ell$-range 140--900 allowing detection of the first 3 acoustic peaks \cite{Halverson02}. DASI was reconfigured to measure CMB polarization, with each antenna measuring a single polarization (left or right circular). In 2002 DASI obtained the first detection of CMB polarization which provided remarkable confirmation of the standard theory of cosmology \cite{Kovac02}.

The CBI was located in the Atacama desert, Chile and operated from 1999--2008 and utilised many of the same components as its ``sister'' instrument, DASI. The CBI consisted of 13 0.9\,m antennas also located on a table that could be moved along 3 axes (az,el,parallactic angle). Like DASI, the antennas did not track giving a constant coverage in the $u,v$ plane, which was improved by rotating the telescope in parallactic angle. The CBI utilised the same 10 channel 10\,GHz digital correlator as was used for DASI. After discussions between the CBI/DASI/VSA groups (circa 2000), the CBI group focussed on high-$\ell$ measurements. The CBI was the first experiment to convincingly show the drop in power due to the Silk damping tail expected from the standard theory at redshifts of $z\sim 1000$ \cite{Pearson03}. The CBI also showed evidence for excess power at $\ell \sim 2000$--3500 \cite{Mason03,Readhead04a}, although this has been debated and has not been confirmed to-date. The CBI was also converted to measuring polarization and detected polarization \cite{Readhead04b}. Later, the CBI was upgraded with larger 1.4\,m dishes to measure clusters via the Sunyaev-Zel'dovich effect \cite{Taylor11}.


\subsection{SZ interferometers}

The study of clusters via the inverse Compton scattering of CMB photons from hot electrons, the Sunyaev-Zel'dovich (SZ) effect, has become an important topic in radio astronomy and cosmology. Interferometers are well-suited to measuring the SZ effect since clusters have a typical angular size of a few arcmin, which requires relatively large single-dish telescopes for ground-based instruments working at $\lesssim 100$\,GHz. One of the earliest measurements of SZ was made with the Ryle interferometer \cite{Jones93}. The VSA, CBI and DASI experiments all made significant contributions to the number of SZ detections. 

The Sunyaev-Zel'dovich Array (SZA) is an 8-element 3.5m dish interferometer located at Owen's Valley Radio Observatory in California. With sensitive receivers operating at 30 and 90\,GHz it has made important contributions to both primary anisotropies at high-$\ell$ \cite{Sharp10} as well as numerous SZ detections including a blind survey \cite{Muchovej11}. The Arcminute MicroKelvin Imager (AMI) is a 15\,GHz experiment located at Lord's Bridge, Cambridge, consisting of eight 3.7\,m dishes, in combination with the larger Ryle telescope dishes for point source subtraction. As well as making targeted observations of known clusters, it is surveying the sky for new clusters. The AMIBA array is a 90\,GHz array located at the Mauna Loa (Haiwaii) with a 10\,m platform. It was originally configured with seven 0.6\,m antennas and then upgraded to 1.2\,m antennas \cite{Liu10}.

\section{The future}

The future for CMB interferometry appears to be limited due to the great need for much higher sensitivity. This is important for both intensity (e.g. for SZ surveys) and, in particular, for polarization where the signal is much weaker. The main goal for CMB experiments is in measuring the large-scale B-modes which require 1000s of detectors. For an interferometer, the number of correlations scales as $n^2$ ($n$ is the number of antennas). The need for sensitivity necessarily leads to very large arrays, possibly with multiple feeds, that would therefore require extremely large correlators (although some correlations could be ignored) operating with very large bandwidths. Although this may be possible with the advent of digital (software) correlators, it is a formidable challenge. One possible exception to this is the bolometer interferometer concept, which essentially works as an adding interferometer (as opposed to multiplying) which gives some (but not all) of the benefits of an interferometer but with the higher sensitivity and larger bandwidths available from bolometers \cite{Piccirillo03,Tucker03}. The QUBIC experiment is aiming to deploy a small array at Antarctica to test this concept \cite{QUBIC11}.

In recent years, there has been great interest in low frequency ($\lesssim 1.4$\,GHz) arrays, aiming towards the Square Kilometre Array (SKA). A major science goal of the SKA is to detect HI during the Epoch of Reionization (EoR) at redshifts $z\gtrsim 6$. This involves measuring tiny ($\lesssim 1$\,mK) fluctuations of the redshifted HI 21\,cm line on scales of $\sim 10$\,arcmin. More recently, the idea of HI intensity mapping \cite{Peterson06} at moderate redshifts has emerged, which potentially could allow detection of Baryon Acoustic Oscillations and thereby constrain dark energy. Both these new exciting areas require similar sensitivities and angular scales as for the CMB primary anisotropies but in the presence of much larger foregrounds and potential systematic errors and calibration. Interferometers may be a good way of achieving this and this is being pursued by a number of groups e.g. \cite{Chen11}.

\acknowledgments

CD acknowledges support from an STFC Advanced Fellowship and an EU Marie-Cure IRG grant under the FP7. CD wishes to thank the following people for reading an early draft and providing useful comments: Keith Grainge, Jean-Christophe Hamilton, Erik Leitch, Tim Pearson, Lucio Piccirillo and Paul Scott.


\begin{thebibliography}{99}

\bibitem{Carlstrom96} Carlstrom, J.~E., Joy, M., \& Grego, L., {\it Interferometric Imaging of the Sunyaev-Zeldovich Effect at 30 GHz}, 1996, \apjl, 456, L75 
\bibitem{Chen11} Chen, X., {\it Radio detection of dark energy - the Tianlai project}, 2011, Scientia Sinica Physica, Mechanica Astronomica, 41, 1358

\bibitem{Dawson02} Dawson, K.~S., Holzapfel, W.~L., Carlstrom, J.~E., et al., {\it Measurement of Arcminute-Scale Cosmic Microwave Background Anisotropy with the Berkeley-Illinois-Maryland Association Array}, 2002, \apj, 581, 86 [{\tt arXiv:astro-ph/0206012}]

\bibitem{Dickinson04} Dickinson, C., Battye, R.~A., Carreira, P., et al., {\it High-sensitivity measurements of the cosmic microwave background power spectrum with the extended Very Small Array}, 2004, \mnras, 353, 732 [{\tt arXiv:astro-ph/0402498}]

\bibitem{Fomalont84} Fomalont, E.~B., Kellermann, K.~I., \& Wall, J.~V., {\it Limits to the small-scale fluctuations in the cosmic background radiation}, 1984, \apjl, 277, L23 

\bibitem{Halverson02} Halverson, N.~W., Leitch, E.~M., Pryke, C., et al., {\it Degree Angular Scale Interferometer First Results: A Measurement of the Cosmic Microwave Background Angular Power Spectrum}, 2002, \apj, 568, 38 [{\tt astro-ph/0104489}]

\bibitem{Harrison00} Harrison, D.~L., Rubi{\~n}o-Martin, J.~A., Melhuish, S.~J., et al., {\it A measurement at the first acoustic peak of the cosmic microwave background with the 33-GHz interferometer}, 2000, \mnras, 316, L24 [{\tt arXiv:astro-ph/0004357}]

\bibitem{Jones93} Jones, M., Saunders, R., Alexander, P., et al., {\it An image of the Sunyaev-Zel'dovich effect}, 1993, \nat, 365, 320

\bibitem{Kovac02} Kovac, J.~M., Leitch, E.~M., Pryke, C., et al., {\it Detection of polarization in the cosmic microwave background using DASI}, 2002, \nat, 420, 772 [{\tt arXiv:astro-ph/0209478}] 

\bibitem{Liu10} Liu, G.-C., Birkinshaw, M., Proty Wu, J.-H., et al.,{\it Contamination of the Central Sunyaev-Zel'Dovich Decrements in AMiBA Galaxy Cluster Observations}, 2010, \apj, 720, 608 [{\tt arXiv:1010.0579}]

\bibitem{Mason03} Mason, B.~S., Pearson, T.~J., Readhead, A.~C.~S., et al., {\it The Anisotropy of the Microwave Background to l = 3500: Deep Field Observations with the Cosmic Background Imager}, 2003, \apj, 591, 540 [{\tt arXiv:astro-ph/0205384}]

\bibitem{Muchovej11} Muchovej, S., Leitch, E., Carlstrom, J.~E., et al., {\it Cosmological Constraints from a 31 GHz Sky Survey with the Sunyaev-Zel'dovich Array}, 2011, \apj, 732, 28 [{\tt arXiv:1012.1610}]

\bibitem{Myers03} Myers, S.~T., Contaldi, C.~R., Bond, J.~R., et al., {\it A Fast Gridded Method for the Estimation of the Power Spectrum of the Cosmic Microwave Background from Interferometer Data with Application to the Cosmic Background Imager}, 2003, \apj, 591, 575 [{\tt arXiv:astro-ph/0205385}]

\bibitem{Padin02} Padin, S., Shepherd, M.~C., Cartwright, J.~K., et al., {\it The Cosmic Background Imager}, 2002, \pasp, 114, 83 [{\tt arXiv:astro-ph/0110124}]

\bibitem{Partridge97} Partridge, R.~B., Richards, E.~A., Fomalont, E.~B., Kellermann, K.~I., \& Windhorst, R.~A., {\it Small-Scale Cosmic Microwave Background Observations at 8.4 GHz}, 1997, \apj, 483, 38

\bibitem{Pearson03} Pearson, T.~J., Mason, B.~S., Readhead, A.~C.~S., et al., {\it The Anisotropy of the Microwave Background to l = 3500: Mosaic Observations with the Cosmic Background Imager}, 2003, \apj, 591, 556 [{\tt arXiv:astro-ph/0205388}]

\bibitem{Peterson06} Peterson, J.~B., Bandura, K., \& Pen, U.~L., {\it The Hubble Sphere Hydrogen Survey}, 2006, Presented at the Moriond Conference (2006) [{\tt arXiv:astro-ph/0606104}] 

\bibitem{Piccirillo03} Piccirillo, L., {\it A large angular scale CMB polarization experiment for Dome-C} 2003, Memorie della Societa Astronomica Italiana Supplementi, 2, 200

\bibitem{QUBIC11} Qubic Collaboration, Battistelli, E., Ba{\'u}, A., et al., {\it QUBIC: The QU bolometric interferometer for cosmology}, 2011, Astroparticle Physics, 34, 705 [{\tt arXiv:1010.0645}]

\bibitem{Radford93} Radford, S.~J.~E., {\it Isotropy of the cosmic background radiation at 3.4 millimeters with 10-arcsec resolution}, 1993, \apjl, 404, L33

\bibitem{Readhead04a} Readhead, A.~C.~S., Mason, B.~S., Contaldi, C.~R., et al., {\it Extended Mosaic Observations with the Cosmic Background Imager}, 2004, \apj, 609, 498 [{\tt arXiv:astro-ph/0402359}]

\bibitem{Readhead04b} Readhead, A.~C.~S., Myers, S.~T., Pearson, T.~J., et al.,{\it Polarization Observations with the Cosmic Background Imager}, 2004, Science, 306, 836 [{\tt arXiv:astro-ph/0409569}]

\bibitem{Sharp10} Sharp, M.~K., Marrone, D.~P., Carlstrom, J.~E., et al., {\it A Measurement of Arcminute Anisotropy in the Cosmic Microwave Background with the Sunyaev-Zel'dovich Array}, 2010, \apj, 713, 82 [{\tt arXiv:0901.4342}]

\bibitem{Scott96} Scott, P.~F., Saunders, R., Pooley, G., et al., {\it Measurements of Structure in the Cosmic Background Radiation with the Cambridge Cosmic Anistropy Telescope}, 1996, \apjl, 461, L1 

\bibitem{Subrahmanyan98} Subrahmanyan, R., Kesteven, M.~J., Ekers, R.~D., Sinclair, M., \& Silk, J., {\it The Australia Telescope search for cosmic microwave background anisotropy}, 1998, \mnras, 298, 1189 [{\tt arXiv:astro-ph/9805245}] 

\bibitem{Taylor11} Taylor, A.~C., Jones, M.~E., Allison, J.~R., et al., {\it The Cosmic Background Imager 2}, 2011, \mnras, 418, 2720 [{\tt arXiv:1108.3950}]

\bibitem{Timbie90} Timbie, P.~T., \& Wilkinson, D.~T., {\it A search for anisotropy in the cosmic microwave radiation at medium angular scales}, 1990, \apj, 353, 140 

\bibitem{Tucker03} Tucker, G.~S., Kim, J., Timbie, P., et al., {\it Bolometric interferometry: the millimeter-wave bolometric interferometer}, 2003, \nar, 47, 1173 

\bibitem{Watson03} Watson, R.~A., Carreira, P., Cleary, K., et al., {\it First results from the Very Small Array - I. Observational methods}, 2003, \mnras, 341, 1057 [{\tt arXiv:astro-ph/0205378}]

\bibitem{White99} White, M., Carlstrom, J.~E., Dragovan, M., \& Holzapfel, W.~L., {\it Interferometric Observation of Cosmic Microwave Background Anisotropies}, 1999, \apj, 514, 12 [{\tt arXiv:astro-ph/9712195}]



\end{thebibliography}
\end{document}